---------------------------------------------------------------
\documentstyle[12pt,amsfonts,amsbsy]{article}
\catcode`\@=11
\long\def\@makefntext#1{ %\parindent 1em
\protect\noindent \hbox to 3.2pt {\hskip-.9pt
$^{{\ninerm\@thefnmark}}$\hfil}#1\hfill} %can be used

\def\thefootnote{\fnsymbol{footnote}}
 \def\@makefnmark{\hbox to 0pt{$^{\@thefnmark}$\hss}}  %original

\def\ps@myheadings{\let\@mkboth\@gobbletwo
\def\@oddhead{\hbox{} %\sl
\rightmark\hfil\ninerm\thepage}
\def\@oddfoot{}\def\@evenhead{\ninerm\thepage\hfil %\sl
\leftmark\hbox{}}\def\@evenfoot{}
\def\sectionmark##1{}\def\subsectionmark##1{}}

\begin{document}

%----------------------------PROCSLA.STY---------------------------------------
\newcommand{\symbolfootnote}{\renewcommand{\thefootnote}
	{\fnsymbol{footnote}}}
\renewcommand{\thefootnote}{\fnsymbol{footnote}}
\newcommand{\alphfootnote}
	{\setcounter{footnote}{0}
	 \renewcommand{\thefootnote}{\sevenrm\alph{footnote}}}

%------------------------------------------------------------------------------
%NEW DEFINED SECTION COMMANDS
\newcounter{sectionc}\newcounter{subsectionc}\newcounter{subsubsectionc}
\renewcommand{\section}[1] {\vspace{0.6cm}\addtocounter{sectionc}{1}
\setcounter{subsectionc}{0}\setcounter{subsubsectionc}{0}\noindent
	{\bf\thesectionc. #1}\par\vspace{0.4cm}}
\renewcommand{\subsection}[1] {\vspace{0.6cm}\addtocounter{subsectionc}{1}
	\setcounter{subsubsectionc}{0}\noindent
	{\it\thesectionc.\thesubsectionc. #1}\par\vspace{0.4cm}}
\renewcommand{\subsubsection}[1]
{\vspace{0.6cm}\addtocounter{subsubsectionc}{1}
	\noindent {\rm\thesectionc.\thesubsectionc.\thesubsubsectionc.
	#1}\par\vspace{0.4cm}}
\newcommand{\nonumsection}[1] {\vspace{0.6cm}\noindent{\bf #1}
	\par\vspace{0.4cm}}

%NEW MACRO TO HANDLE APPENDICES
\newcounter{appendixc}
\newcounter{subappendixc}[appendixc]
\newcounter{subsubappendixc}[subappendixc]
\renewcommand{\thesubappendixc}{\Alph{appendixc}.\arabic{subappendixc}}
\renewcommand{\thesubsubappendixc}
	{\Alph{appendixc}.\arabic{subappendixc}.\arabic{subsubappendixc}}

\renewcommand{\appendix}[1] {\vspace{0.6cm}
        \refstepcounter{appendixc}
        \setcounter{figure}{0}
        \setcounter{table}{0}
        \setcounter{equation}{0}
        \renewcommand{\thefigure}{\Alph{appendixc}.\arabic{figure}}
        \renewcommand{\thetable}{\Alph{appendixc}.\arabic{table}}
        \renewcommand{\theappendixc}{\Alph{appendixc}}
        \renewcommand{\theequation}{\Alph{appendixc}.\arabic{equation}}
%       \noindent{\bf Appendix \theappendixc. #1}\par\vspace{0.4cm}}
        \noindent{\bf Appendix \theappendixc #1}\par\vspace{0.4cm}}
\newcommand{\subappendix}[1] {\vspace{0.6cm}
        \refstepcounter{subappendixc}
        \noindent{\bf Appendix \thesubappendixc. #1}\par\vspace{0.4cm}}
\newcommand{\subsubappendix}[1] {\vspace{0.6cm}
        \refstepcounter{subsubappendixc}
        \noindent{\it Appendix \thesubsubappendixc. #1}
	\par\vspace{0.4cm}}

%------------------------------------------------------------------------------
%MARCO FOR ABSTRACT BLOCK
\def\abstracts#1{{
	\centering{\begin{minipage}{30pc}\tenrm\baselineskip=12pt\noindent
	\centerline{\tenrm ABSTRACT}\vspace{0.3cm}
	\parindent=0pt #1
	\end{minipage} }\par}}

%------------------------------------------------------------------------------
%NEW MACRO FOR BIBLIOGRAPHY
\newcommand{\bibit}{\it}
\newcommand{\bibbf}{\bf}
\renewenvironment{thebibliography}[1]
	{\begin{list}{\arabic{enumi}.}
	{\usecounter{enumi}\setlength{\parsep}{0pt}
%1.25cm IS STRICTLY FOR PROCSLA.TEX ONLY
\setlength{\leftmargin 1.25cm}{\rightmargin 0pt}
%0.52cm IS FOR NEW DATA FILES
%\setlength{\leftmargin 0.52cm}{\rightmargin 0pt}
	 \setlength{\itemsep}{0pt} \settowidth
	{\labelwidth}{#1.}\sloppy}}{\end{list}}

%------------------------------------------------------------------------------
%FOLLOWING THREE COMMANDS ARE FOR 'LIST' COMMAND.
\topsep=0in\parsep=0in\itemsep=0in
\parindent=1.5pc

%LIST ENVIRONMENTS
\newcounter{itemlistc}
\newcounter{romanlistc}
\newcounter{alphlistc}
\newcounter{arabiclistc}
\newenvironment{itemlist}
    	{\setcounter{itemlistc}{0}
	 \begin{list}{$\bullet$}
	{\usecounter{itemlistc}
	 \setlength{\parsep}{0pt}
	 \setlength{\itemsep}{0pt}}}{\end{list}}

\newenvironment{romanlist}
	{\setcounter{romanlistc}{0}
	 \begin{list}{$($\roman{romanlistc}$)$}
	{\usecounter{romanlistc}
	 \setlength{\parsep}{0pt}
	 \setlength{\itemsep}{0pt}}}{\end{list}}

\newenvironment{alphlist}
	{\setcounter{alphlistc}{0}
	 \begin{list}{$($\alph{alphlistc}$)$}
	{\usecounter{alphlistc}
	 \setlength{\parsep}{0pt}
	 \setlength{\itemsep}{0pt}}}{\end{list}}

\newenvironment{arabiclist}
	{\setcounter{arabiclistc}{0}
	 \begin{list}{\arabic{arabiclistc}}
	{\usecounter{arabiclistc}
	 \setlength{\parsep}{0pt}
	 \setlength{\itemsep}{0pt}}}{\end{list}}

%------------------------------------------------------------------------------
%FIGURE CAPTION
\newcommand{\fcaption}[1]{
        \refstepcounter{figure}
        \setbox\@tempboxa = \hbox{\tenrm Fig.~\thefigure. #1}
        \ifdim \wd\@tempboxa > 6in
           {\begin{center}
        \parbox{6in}{\tenrm\baselineskip=12pt Fig.~\thefigure. #1 }
            \end{center}}
        \else
             {\begin{center}
             {\tenrm Fig.~\thefigure. #1}
              \end{center}}
        \fi}

%TABLE CAPTION
\newcommand{\tcaption}[1]{
        \refstepcounter{table}
        \setbox\@tempboxa = \hbox{\tenrm Table~\thetable. #1}
        \ifdim \wd\@tempboxa > 6in
           {\begin{center}
        \parbox{6in}{\tenrm\baselineskip=12pt Table~\thetable. #1 }
            \end{center}}
        \else
             {\begin{center}
             {\tenrm Table~\thetable. #1}
              \end{center}}
        \fi}

%------------------------------------------------------------------------------
%ACKNOWLEDGEMENT: this portion is from John Hershberger
\def\@citex[#1]#2{\if@filesw\immediate\write\@auxout
	{\string\citation{#2}}\fi
\def\@citea{}\@cite{\@for\@citeb:=#2\do
	{\@citea\def\@citea{,}\@ifundefined
	{b@\@citeb}{{\boldsymbol ?}\@warning
	{Citation `\@citeb' on page \thepage \space undefined}}
	{\csname b@\@citeb\endcsname}}}{#1}}

\newif\if@cghi
\def\cite{\@cghitrue\@ifnextchar [{\@tempswatrue
	\@citex}{\@tempswafalse\@citex[]}}
\def\citelow{\@cghifalse\@ifnextchar [{\@tempswatrue
	\@citex}{\@tempswafalse\@citex[]}}
\def\@cite#1#2{{$\null^{#1}$\if@tempswa\typeout
	{IJCGA warning: optional citation argument
	ignored: `#2'} \fi}}
\newcommand{\citeup}{\cite}

%------------------------------------------------------------------------------
%FOR FNSYMBOL FOOTNOTE AND ALPH{FOOTNOTE}
\def\fnm#1{$^{\mbox{\scriptsize #1}}$}
\def\fnt#1#2{\footnotetext{\kern-.3em
	{$^{\mbox{\sevenrm #1}}$}{#2}}}

%------------------------------------------------------------------------------
\font\twelvebf=cmbx10 scaled\magstep 1
\font\twelverm=cmr10 scaled\magstep 1
\font\twelveit=cmti10 scaled\magstep 1
\font\elevenbfit=cmbxti10 scaled\magstephalf
\font\elevenbf=cmbx10 scaled\magstephalf
\font\elevenrm=cmr10 scaled\magstephalf
\font\elevenit=cmti10 scaled\magstephalf
\font\bfit=cmbxti10
\font\tenbf=cmbx10
\font\tenrm=cmr10
\font\tenit=cmti10
\font\ninebf=cmbx9
\font\ninerm=cmr9
\font\nineit=cmti9
\font\eightbf=cmbx8
\font\eightrm=cmr8
\font\eightit=cmti8
%%%%%%%%%%%%%%%%%%%%%%%%%%
\def\fmref#1{(\ref{#1})}
%%%%%%%%%%%%%%%%%%%%%%%%%%%%%%%%%%%%%%%%%%%%%%%%%%%%%%%%%%%%%%%%%%%%%%%%%
%
%---- Fock space and Liouville space state vectors
\def\lds{|\mkern-2.5mu|}
\def\rds{|\mkern-2.5mu|}
\def\ldb{\langle\mkern-4mu\langle}
\def\rdb{\rangle\mkern-4mu\rangle}
\def\ldc{(\mkern-4mu(}
\def\rdc{)\mkern-4mu)}
\def\RGV{\lds W^R \rdc}
\def\LGV{\ldc W^L \rds}
\def\RTV{\lds W^R_\beta \rdc}
\def\LTV{\ldc W^L_\beta \rds}
\def\Av#1{\left\langle #1\vphantom{\int}\right\rangle}
%---- creation and annihilation operators operators
\def\dag{{\setlength{\unitlength}{.09mm}
              \picture(18,30)
              \put(2,20){\line(1,0){16}}
              \put(10,0){\line(0,1){30}}
              \endpicture}}
\def\ankh{{\setlength{\unitlength}{.09mm}
              \picture(18,30)
              \put(2,23){\line(1,0){16}}
              \put(2,17){\line(1,0){16}}
              \put(7,0){\line(0,1){30}}
              \put(13,0){\line(0,1){30}}
              \endpicture}}
\def\wtilde#1{\widetilde{#1}}
\def\what#1{\widehat{#1}}
\def\a{a}
\def\at{\wtilde{a}}
\def\ad{a^\dagger}
\def\atd{\wtilde{a}^\dagger}
%
%---- General abbreviations
%
\def\vec#1{\boldsymbol{#1}}
\def\i{\mathrm{i}}
\def\r{\boldsymbol{r}}
\def\e{\mathop{\mathrm{e}}\nolimits}
\def\ee#1{\e^{#1}}
\makeatletter
\baselineskip=16pt
\centerline{\tenbf THERMO FIELD DYNAMICS FOR QUANTUM FIELDS }
\centerline{\tenbf WITH CONTINUOUS MASS SPECTRUM }
\centerline{\tenbf APPLIED TO NUCLEAR PHYSICS\footnote{Work supported by GSI}
\footnote{Talk given at the ``3rd Workshop on Thermal Field Theories and their
applications'', Banff 1993. To appear in the proceedings
(World Scientific, 1994).}}
\vspace{0.8cm}
\centerline{\tenrm PETER A. HENNING\footnote{
  E-mail: phenning@tpri6c.gsi.de}}
\baselineskip=13pt
\centerline{\tenit Institut f. Kernphysik, TH Darmstadt, and}
\centerline{\tenit Gesellschaft f. Schwerionenforschung GSI}
\centerline{\tenit P.O.Box 110552, D-64220 Darmstadt}
\vspace{0.9cm}
\abstracts{Transport coefficients are obtained
by incorporating a gauge principle
into thermo field dynamics of inhomogeneous systems.
In contrast to previous derivations, neither imaginary time
arguments nor perturbation theory in powers of a coupling constant
are used in the calculation. Numerical values are calculated for
the pion component in hot nuclear matter.}
\vfil
%\vspace{0.8cm}
\twelverm   %modified by CLee 23/07/93
\baselineskip=14pt
%%%%%%%%%%%%%%%%%%%%%%%%%%%%%%%%%%%%%%%%%%%%%%%%%%%%%%%%%%%%%%%%%%%%%%%%%%
\section{Introduction}
In the presence of homogeneous
matter, or in a heat bath, the irreducible
representations of the space-time symmetry group are characterized
by a continuous energy parameter rather than by a mass-shell
constraint. It follows, that physical ``particles'' are unstable
and therefore a naive perturbation theory starting from free on-shell
particles may lead to inconsistencies. To circumvent this
problem, a perturbative expansion in terms of generalized free
fields may be defined, cf. ref. \cite{L88} and papers quoted there.

Thermo field dynamics (TFD, see ref. \cite{Ubook})
is a necessary input to this solution, since in contrast to the
Schwinger-Keldysh (closed-time-path, CTP) method \cite{SKM}
of quantum statistical mechanics, it possesses two different
(anti-)commuting representations of the underlying field algebra.
Apart from this mathematical aspect however, TFD also contains
a concept which extrexemly simplifies its practical application:
Physical excitations are obtained as superposition
of {\em thermal quasi-particle} states. Due to the facts outlined above,
this must be a continuous superposition, with a weight function
that is nothing but the {\em spectral function\/} of a field theory.

To establish the notation, we first discuss TFD for a single bosonic
oscillator, introducing canonical creation and annihilation operators
$\ad$, $\a$, $\atd$, $\at$ for the two representations.
The $\a$, $\at$ operators annihilate
the physical vacuum, and the two sets
are transformed into another by means of an anti-unitary
mapping, called the {\it tilde} conjugation \cite{Ubook}.
One may construct these representations in
the Liouville space of the oscillator, where the generator of
time evolution is
\begin{equation}\label{nn}
\widehat{H} = \omega\,\widehat{N}
,\;\;\;\;\;\widehat{N} = \ad\a-\atd\at
\;.\end{equation}
This operator possesses a continuous symmetry, since it is
invariant under Bogoliubov transformations to a new set of operators
\begin{equation}\label{bdef}
        \left(\array{r}\xi\\
\wtilde{\xi}^\ankh\endarray\right)=
  {\cal B}
  \left(\array{r}\a \\ \atd\endarray\right)
 \;\;\;\;
\left(\array{r}\xi^\ankh\\ -\wtilde{\xi}\endarray\right)^T=
  \left(\array{r}\ad\\-\at\endarray\right)^T
  {\cal B}^{-1}
\;,\end{equation}
where ${\cal B}$ is a (real) 2$\times$2 matrix with determinant 1.
These matrices form the symplectic
group in two dimensions Sp(2) = SU(1,1), it is a noncompact
and non-abelian group \cite{L86}.

In thermo field dynamics, the new set of operators obeys
thermal state conditions:
\begin{equation}\label{tsc}
\xi\RGV=0\;,\;\;\;\wtilde{\xi}\RGV=0 \;,\;\;\;
\LGV\xi^\ankh=0 \;,\;\;\;\LGV\wtilde{\xi}^\ankh=0
\;, \end{equation}
where $\LGV$ and $\RGV$ are left and right thermal state.
It can be shown, that a particularly simple parameterization
of TFD is obtained with
\begin{equation}\label{bdef2}
\RGV =\lds W\rdb\;,\;\;
\;\;\;\LGV =\ldb 1 \rds\;,\;\;\;
{\cal B} =
\left(\array{cc}1+ n & -n\\
                -1 & 1\endarray\right)
\;.\end{equation}
Here, $\lds 1 \rdb$ is the Liouville vector associated with the
unit operator, and $\lds W \rdb$ is the Liouville vector associated
with the density matrix $W$ of the quantum system. Statistical averages
for an operator ${\cal E}$ are calculated as $\ldb 1 \rds {\cal E}\otimes 1
\lds W \rdb$, i.e., as a matrix element instead of a trace over the
Hilbert space of the oscillator.

%%%%%%%%%%%%%%%%%%%%%%%%%%%%%%%%%%%%%%%%%%%%%%%%%%%%%%%%%%%%%%%%%%%
\section{Gauging the Bogoliubov symmetry}
The Bogoliubov symmetry discussed above
is a {\em global\/} symmetry, which for the
simple case of a single oscillator means that it is
time-independent. We now consider the question of a {\em local\/}
Bogoliubov symmetry, which may arise in case the transformation
matrix depends on some external parameters \cite{h90ber}.
For convenience, we consider these external parameters as a vector
$\r$ in some abstract space.

Requiring the locality of a former global symmetry
is called, in accordance with
the methods applied to fundamental forces of nature,
the {\em gauging\/} of the Bogoliubov symmetry.
This name however is more than a simple analogy:
It is known, that even in simple
dynamical systems structures arise, that most closely resemble
the coupling of the system to an external gauge potential
\cite{WZ84}.

The physical particle operators depend
on the parameter vector $\r$. Hence, the former will change
with time, for infinitesimal changes of $\r$ according to
\begin{eqnarray}\label{pdef}
\r \;\longrightarrow\;\r + \delta\r:&
  \left(\array{c}\a(t)\\ \atd(t)\endarray\right)
  &\longrightarrow \;\left(1 + {\cal P}(\r)\delta\r\right)\;
  \left(\array{c}\a(t)\\ \atd(t)\endarray\right)\nonumber \\
 &\left(\array{c}\ad(t)\\ -\at(t)\endarray\right)^T
  &\longrightarrow \;
  \left(\array{c}\ad(t)\\ -\at(t)\endarray\right)^T\;
  \left(1 - {\cal P}(\r)\delta\r\right)
\;.\end{eqnarray}
${\cal P}$ is a vector in
$\r$-space, ${\cal P}=({\cal P}^1,{\cal P}^2,\dots)$, each
component forming a $2\times2$-matrix.

This implies, that the change of the physical
particles with the external parameters has a generator
\begin{equation}\label{qdf}
\widehat{\cal Q} = \i
   \left(\array{c} \ad(t)\\ -\sigma\at(t)\endarray\right)^T
   \,{\cal P}(\r)\,\dot{\r}\,
   \left(\array{c} \a(t)\\ \atd(t)\endarray\right)
\;.\end{equation}
The total time evolution of the system is generated by the
sum of $\widehat{H}$ and $\widehat{\cal Q}$, i.e.,
the Liouville equation for an operator ${\cal E}$ reads
\begin{equation}\label{cvv}
\i\frac{\partial}{\partial t}\, \lds {\cal E}(t) \rdb  =
            \left( \widehat{H}+ \widehat{\cal Q}\right)\,
                \lds {\cal E}(t)\rdb
\;.\end{equation}
Now consider the thermal quasi-particle operators,
which under an $\r$-evolution change with a {\em different\/}
$\widehat{\cal Q}$.
{}From eqns. \fmref{bdef} and
\fmref{pdef} follows, that it is obtained as
\begin{equation}\label{pmod}
   {\cal P}^\prime(\r) =
  {\cal B}(\r)\,{\cal P}(\r)\,{\cal B}^{-1}(\r)+
   {\cal B}(\r)\,\left(\nabla_{\r}{\cal B}^{-1}(\r)\right)
\;.\end{equation}
Hence, under a change of the
operator description mediated by a Bogoliubov transformation,
${\cal P}$ transforms
inhomogeneously like a gauge potential.  The
substitution of the $\xi$-operators for the $\a$-operators is
therefore, in this new language, a {\em gauge transformation\/}.
Finally, we require diagonality of the hamiltonian
for thermal quasi-particles, i.e.,
\begin{equation}\label{dix}
\left(\widehat{H}+\widehat{\cal Q}\right)[\xi] =
        \omega \left(\xi^\ankh\xi-\wtilde{\xi}^\ankh\wtilde{\xi}\right)
\;.\end{equation}
Returning to eqn. \fmref{pmod}, this means that
${\cal P}^\prime$ has to vanish. This requirement fixes the
connection between the physical particles and the statistical
quasi-particles unambiguously: The change of the physical
particle operators with the external parameters
according to  \fmref{pdef} has a gauge potential
given by
\begin{equation}\label{gpo}
{\cal P}(\r)=  - \left(\nabla_r\,{\cal B}^{-1}(\r)\right){\cal B}(\r)
\;.\end{equation}
To summarize these considerations: {\em Gauging\/} the Bogoliubov
symmetry of the time evolution in Liouville space, while keeping
the thermal quasi-particle operators fixed, introduces a coupling
of the physical particle representation to gradients in the external
parameters.

%%%%%%%%%%%%%%%%%%%%%%%%%%%%%%%%%%%%%%%%%%%%%%%%%%%%%%%%%%%%%%%%%%
\section{Interacting fields with local Bogoliubov symmetry}
An interacting charged scalar field at a fixed time can be expanded into
momentum eigenmodes as
\begin{equation}\label{cbf1}
\phi_x  =
   \int\!\! \frac{d^3\vec{k}}{\sqrt{(2\pi)^3}}
   \left( a^\dag_{k-}(t)\,\ee{-\i \vec{k}\vec{x}} +
          a_{k+}(t)\,        \ee{ \i \vec{k}\vec{x}}\right)
\;,\end{equation}
and similarly for the tilde-field.
$\vec{k}$ is the three-momentum of the modes, and in this notation
$a^\dag_{k-}(t)$ creates a negatively charged excitation with momentum
$\vec{k}$, while $a_{k+}(t)$ annihilates a positive charge.
We will henceforth distinguish the two different charges
by an additional index $l=\pm$ whenever possible.

In general, the commutation relations of the $a$-operators are
not known: They can be very complicated due to the interaction.
However, for the purpose of calculating bilinear expectation
values, e.g. propagators, it is sufficient to know the expectation
value of the commutator functions. Hence, for this purpose, the
scalar field operators may be considered as generalized free
fields, with an expansion into modes with definite energy and momentum
that are created and annihilated by a new set of operators.
The details of the expansion procedure in general
are outlined in refs. \cite{L88,hu92a,habil}, its generalization to
inhomogeneous states in \cite{NUY92,h93trans}. Here it is sufficient
to write down the generalization of eqn. \fmref{bdef}
\begin{eqnarray}\label{bbg5}
&&\left({\array{r} a_{kl}(t)\\
          \widetilde{a}^\dag_{kl}(t)\endarray}\right)\;=
\nonumber  \\
&& \displaystyle\int\limits_0^\infty\!\!dE\,\int\!\!d^3\vec{q}
  \;\left({\cal A}_l(E,(\vec{q}+\vec{k})/2)\right)^{1/2}\,
  \left(\widetilde{\cal B}^{-1}_l(E,\vec{q},\vec{k})\right)^\star
  \,\left({\array{r}\xi_{Eql}\\
              \widetilde{\xi}^\ankh_{Eql}\endarray}\right)
  \,\e^{-\i  Et}
 \;.\end{eqnarray}
${\cal A}_l(E,\vec{k})$ is a positive weight function
with support on the positive energy axis, and the transformation matrix
has the form
\begin{equation}\label{gb}
\widetilde{\cal B}_l(E,\vec{q},\vec{k}) = \left( { \array{lr}
   \left(\delta^3(\vec{q}-\vec{k}) + N_l(E,\vec{q},\vec{k})\right)
            \;\;\;& -N_l(E,\vec{q},\vec{k}) \\
   -\delta^3(\vec{q}-\vec{k})     & \delta^3(\vec{q}-\vec{k})
   \endarray} \right)
\;\end{equation}
similar to the matrix in eqn. \fmref{bdef2}.
The Bogoliubov parameter $N_l$ appearing here is the Fourier
transform of a space-local quantity
\begin{equation}\label{nloc}
N_l(E,\vec{q},\vec{k}) = \frac{1}{(2\pi)^3}\;
        \int\!\!d^3\vec{z}\,\e^{-\i
  (\vec{q}-\vec{k})\vec{z} }\,
  n_l(E,(\vec{q}+\vec{k})/2,\vec{z})
\;.\end{equation}
The thermal state conditions are similar to eqn. \fmref{bdef2},
but hold for both charges and for each value of $E,\vec{k}$.
The time derivative of the $a$-operators
in the Heisenberg picture then is
\begin{eqnarray}\label{tide}
 \i \frac{\partial}{\partial t}\,
 \left({\array{r} a_{kl}(t)\\
         \widetilde{a}^\dag_{kl}(t)\endarray}\right)&&=
 \Omega_{kl}\,\left({\array{r} a_{kl}(t)\\
         \widetilde{a}^\dag_{kl}(t)\endarray}\right) \nonumber \\
 +&&\int\limits_0^\infty\!\!dE\int\!\!d^3\vec{q}
  \,\frac{\left({\cal A}_l(E,(\vec{k}+\vec{q})/2)\right)^{1/2}}{Z_{ql}}
  \;E\,N_l(E,\vec{k},\vec{q})\,\times \nonumber \\
&&  \left(\left({\cal A}_l(E,\vec{k})\right)^{1/2}-
        \left({\cal A}_l(E,\vec{q})\right)^{1/2}\right)\,
\left({\array{lr} 1 & 1\\ 1 & 1 \endarray}\right)\,
  \left({\array{r}a_{ql}(t)\\
           \widetilde{a}^\dag_{ql}(t)\endarray}\right)
\;,\end{eqnarray}
which we use to determine the operator $\widehat{{\cal Q}}$.
The $Z$-factor is defined as
\begin{equation}\label{norm2}
Z_{kl}  = \int\limits_0^\infty\!\!dE\,
        {\cal A}_l(E,\vec{k})
\;,\end{equation}
and $\Omega_{kl}=1/(2 Z_{kl})$.
Obviously the additional term
in the time evolution vanishes for translationally invariant
$n_l$, i.e., when $N_l(E,\vec{k},\vec{q})$ is proportional to
$\delta^3(\vec{k}-\vec{q})$. This is consistent with eqn. \fmref{gpo},
henceforth the generator $\widehat{{\cal Q}}$ is
\begin{equation}\label{hen}
\widehat{{\cal Q}} = \sum\limits_{l=\pm}
\int\!\!d^3\vec{k}\,d^3\vec{q}\,
        \left({\array{r} a^\dag_{kl}(t)\\
         -\widetilde{a}_{kl}(t)\endarray}\right)^T\,
\left({\array{lr} 1 & 1\\ 1 & 1 \endarray}\right)\,
        {\cal P}_l(\vec{k},\vec{q})\,
        \left({\array{r} a_{ql}(t)\\
         \widetilde{a}^\dag_{ql}(t)\endarray}\right)\,
\;.\end{equation}
It has a kernel
\begin{eqnarray}\label{ke}
{\cal P}_l(\vec{k},\vec{q})
        &=&\int\limits_0^\infty\!\!dE
  \,\left\{\frac{\left(
          {\cal A}_l(E,(\vec{k}+\vec{k})/2)\right)^{1/2}}{
                 Z_{kl}\,Z_{ql}}\right\}
  \,E\,N_l(E,\vec{k},\vec{k}) \nonumber \\
  &&\;\;\;\;\;\;\;\;\;\;\left(\left({\cal A}_l(E,\vec{k})\right)^{1/2}-
        \left({\cal A}_l(E,\vec{k})\right)^{1/2}\right)
\;.\end{eqnarray}
The dimension of the numerical factors in this kernel
is mass$^2$, due to the energy normalization factors contained in the
$a$-operators for the interacting field (\ref{cbf1}).

%%%%%%%%%%%%%%%%%%%%%%%%%%%%%%%%%%%%%%%%%%%%%%%%%%%%%%%%%%%%%%%%%%%%%%%%%%
\section{Transport coefficients}
Observables are expressed as functionals of the interacting fields
-- and therefore as functionals of the operators $a$, $a^\dag$.
Hence, although the state we consider is stationary in terms of
the basis defined by the $\xi$-operators, momentum mixing as
introduced above will result in a non-trivial time evolution
of {\em physical\/} quantities.
Two quantities we are interested in are the
particle current operator of the complex scalar field
and its expectation value,
\begin{equation}
\widehat{\vec{j}}(t,\vec{x}) = \i \,
   \left(\phi^\star_x\nabla\phi_x - \phi_x\nabla\phi^\star_x
   \right)\;,\;\;\;
   \vec{j}(t,\vec{x})=
    \delta\Av{\widehat{\vec{j}}(t,\vec{x})}
\;\end{equation}
and the energy current
\begin{equation}
\widehat{\vec{E}}(t,\vec{x}) =
   \left(\partial_t\phi^\star_x\nabla\phi_x +
   \partial_t\phi_x\nabla\phi^\star_x
   \right)\;,\;\;\;
   \vec{E}(t,\vec{x})=
    \delta\Av{\widehat{\vec{E}}(t,\vec{x})}
\;.\end{equation}
The response of such a current to the gradients is, to first order
in the gradients, given by
\begin{equation}\label{td}
\delta\Av{\vec{j}(t,\vec{x})} = \i \int\limits_{t_0}^t\!\!d\tau
  \Av{\left[\vec{j}(t-\tau,\vec{x}),\widehat{\cal Q}\right]}
\;.\end{equation}
The commutator contains {\em four\/} field
operators instead of {\it two\/}. The expectation value of this
commutator therefore cannot be expressed completely through the commutator
function of generalized free fields. Hence in the following,
we neglect higher order correlations and use the canonical
commutation relations of the $\xi$-operators to obtain in the
stationary limit
\begin{eqnarray}
&&\vec{j}_l(t,\vec{x}) = \nonumber \\ && \;\;\;-2\pi \i \,
\int\limits_0^\infty\!\!dE\,
\int\!\!\frac{d^3\vec{k}\,d^3\vec{k}^\prime}{(2\pi)^3}\,
        {\cal A}_l(E,\vec{k})\,{\cal A}_l(E,\vec{k}^\prime)
        \,\ee{\i (\vec{k}-\vec{k}^\prime)\vec{x}}
        \,\left(\vec{k}+\vec{k}^\prime\right)
        \,{\cal P}_l(\vec{k},\vec{k}^\prime)
\;\end{eqnarray}
for each species. The energy current is formally quite similar,
and the total currents are for both cases
\begin{eqnarray}
\vec{j}(t,\vec{x}) &=&
\vec{j}_+(t,\vec{x})-\vec{j}_-(t,\vec{x}) \nonumber  \\
\vec{E}(t,\vec{x})&=&
\vec{E}_+(t,\vec{x})+\vec{E}_-(t,\vec{x})
\;.\end{eqnarray}
In principle the above expressions can be calculated, when
the spectral functions and the space-dependence of $n_l$ are
given. However, it is better justified to perform a gradient
expansion for the occupation number density $n_l$, since the
usual spectral function is correct also only to first order
in these gradients.  The $i$-th vector component of the $l$-charged currents
generated by the inhomogeneity of the system is then
\begin{eqnarray}\label{cur}
\vec{j}^{(i)}_l(t,\vec{x})&=\;\displaystyle
   2\pi\,\int\!\!\frac{d^3\vec{k}}{(2\pi)^3}\,
   \frac{\vec{k}^{(i)}}{2\,Z^2_{Ql}}&
   \int\!\!dE\,\left({\cal A}_l(E,\vec{k})\right)^2
   \times\nonumber \\
 &&\int\!\!dE^\prime\,E^\prime\,\left\{
   \frac{\partial n_l(E^\prime,\vec{k},\vec{x})}{\partial \vec{x}^{(j)}}
   \frac{\partial {\cal A}_l(E^\prime,\vec{k})}{\partial \vec{k}^{(j)}}
   \right\}  \\
\vec{E}^{(i)}_l(t,\vec{x})&=\;\displaystyle
   2\pi\,\int\!\!\frac{d^3\vec{k}}{(2\pi)^3}\,
   \frac{\vec{k}^{(i)}}{2\,Z^2_{Ql}}&
   \int\!\!dE\,E\,\left({\cal A}_l(E,\vec{k})\right)^2
   \times \nonumber  \\
 &&\int\!\!dE^\prime\,E^\prime\,\left\{
   \frac{\partial n_l(E^\prime,\vec{k},\vec{x})}{\partial \vec{x}^{(j)}}
   \frac{\partial {\cal A}_l(E^\prime,\vec{k})}{\partial \vec{k}^{(j)}}
   \right\}
\;.\end{eqnarray}
Before this is evaluated further, note that the expression
in the curly brackets is nothing but the first term of a full
gradient expansion of the product of $n_l$ and ${\cal A}_l$, i.e.,
the Poisson bracket.
It is therefore clear, that the above expression
is useful also for situations, where the space-dependence of
${\cal A}_l$ cannot be neglected: the Poisson bracket then also has
a contribution with a space-derivative acting on ${\cal A}_l$ and a
momentum derivative acting on $n_l$.
The above expression is also interesting in view of the
original equation (\ref{td}), since the Poisson bracket
is the analogon of the commutator, i.e., it contributes
a factor $\hbar$ to the current.

Using as local equilibrium spectral function the one for global
equilibrium (correct to first order in the gradients, cf. \cite{habil}),
and as distribution function
\begin{equation}\label{nb2}
n_\pm(E,\vec{k},\vec{x}) =
\frac{f_\pm(E,\vec{x})}{1-f_\pm(E,\vec{x})} =
\frac{1}{\exp\left(\beta(\vec{x}) (E\mp\mu(\vec{x}))\right)-1}
\;,\end{equation}
the derivative with respect to $\vec{x}$ has two components:
Along the temperature gradient and along the gradient of $\mu$.

The same decomposition into $T$ and $\mu$-gradient terms
then holds for the currents, i.e.,
with transport coefficients $L_{ij}$ we obtain
\begin{eqnarray}\label{coef}
\vec{j}(t,\vec{x})& =
- L_{11} \nabla\mu &- L_{12} \frac{\nabla T}{T}
   \nonumber \\
\vec{E}(t,\vec{x})& =
- L_{21} \nabla\mu &- L_{22} \frac{\nabla T}{T}
\;.\end{eqnarray}
Note, that $L_{12}$ and $L_{21}$ can be different,
i.e. the Onsager relation $L_{ij}=L_{ji}$ is not necessarily fulfilled.
The reason for this is clear: in contrast to ordinary many-body quantum
physics, the present formulation exhibits dissipation already on the
tree-graph level \cite{L88}. In other words,
the formulation of an interacting
theory with continuous spectral functions is {\em not}
micro-reversible, physical states have a finite lifetime.
However, the {\em complete\/} system, including all species of
particles and interactions, by definition has a single
partition function. Hence, the {\em global\/}
Onsager relations are fulfilled.

{}From the above currents, one obtains the
thermal conductivity for the interacting scalar field
by subtracting the convective part
\glossary{$\lambda$ \>\' thermal conductivity}
\glossary{$d$ \>\' diffusion coefficient}
\begin{equation}
\lambda = \frac{1}{T}\,\left( L_{22} -\frac{L_{12}L_{21}}{L_{11}}\right)
\;.\end{equation}
$L_{11}=d$ is the diffusion coefficient
(or mobility).
Before considering numerical results obtained with these equations,
let us study the case of a simple spectral function
\begin{equation}\label{simple}
{\cal A}_+(E,\vec{k}) = {\cal A}_-(E,\vec{k})
= \frac{2E\gamma_k}{\pi}\,
   \frac{1}{(E^2-\Omega^2_k)^2+4 E^2 \gamma^2_k}
\;\end{equation}
with $\Omega_k^2=\epsilon_k^2+\gamma_k^2$ and
$\gamma_k\ll\epsilon_k$.
In the absence of a chemical potential, $L_{12}$ and
$L_{21}$ are zero, but diffusion coefficient and
thermal conductivity are, to lowest order in the width
for $\mu=0$,
\begin{eqnarray}\label{lam1}
d_0& =& -\frac{1}{T}  \,\int\!\!\frac{d^3\vec{k}}{(2\pi)^3}\,
   \frac{ 1 }{3 \gamma_k}\,\frac{\partial}{\partial k}\,
        \epsilon_k\,n(\epsilon_k)\left(1+n(\epsilon_k)\right)
 \nonumber \\
\lambda_0& =& -\frac{1}{T}  \,\int\!\!\frac{d^3\vec{k}}{(2\pi)^3}\,
   \frac{ k\,\epsilon_k }{3 \gamma_k}\,\frac{\partial}{\partial k}\,
        \epsilon_k\,n(\epsilon_k)\left(1+n(\epsilon_k)\right)
\;.\end{eqnarray}
Here, $n(\epsilon_k)$ is the local Bose function taken at energy
$\epsilon_k$. Apart from the momentum factors, a similar representation
for $\lambda_0$ was obtained in ref. \cite{HST84}. The
difference can be attributed to the fact that
in this reference a hydrodynamical
picture was assumed together with $\beta\gamma_k\ll 1$.
We thus make a comparison in the high temperature limit, where
one obtains
\begin{equation}\label{asy}
   \lim_{T\rightarrow\infty} \lambda_0 = \frac{A}{\bar{\gamma}}\,T^3
\;.\end{equation}
Here, $\bar{\gamma}$ is some momentum averaged width and $A$
a numerical factor. The same limiting behavior is obtained in
\cite{HST84}.

The above expressions are also in accordance
with naive expectations: The thermal conductivity and
diffusion coefficient diverge, if
the width of the particles is reduced, i.e., if the interaction
is removed. In this case the time needed for the relaxation
of a temperature disturbance is infinite.

As stated above, The Poisson bracket contributes a factor $\hbar$,
while in a naive counting
the inverse width contributes a factor $1/\hbar$ to the
transport coefficient. This cancellation
seems to be the main source for the difficulties one has
in obtaining kinetic coefficients from quantum field theory.

%%%%%%%%%%%%%%%%%%%%%%%%%%%%%%%%%%%%%%%%%%%%%%%%%%%%%%%%%%%%%%%%%%%%%%%%%%%%%%%
\section{Pionic transport properties}\label{transc}
In the next step, we apply the model to
pions in a homogeneous gas of nucleons and
$\Delta_{33}$-resonances. Instead of the above simplistic form,
the more realistic pionic spectral function from ref. \cite{h93pion} is
used. It exhibits a strong mixing of the pion with a collective
exciation of $\Delta$-particle/nucleon-hole pairs, hence has a
richer structure with two resonance branches.
In the figure, the product $\lambda T$ and and the diffusion
coefficient $d$ are plotted for this spectral function.
%%%%%%%%%%%%%%%%%%%%%%%%%%%%%%%%%%%%%%%%%%%%%%%%%%%
\begin{figure}
\vspace*{95mm}
%%
%% This is for dvips (IBM)
\includegraphics{lambda.ps}
%% This is for specialhost (AMIGA)
%%\special{ifffile="lambda.iff" hoffset=0pt voffset=0pt
%%         mode=bw hsize=150mm vsize=235mm}
%%
\vspace*{95mm}
%%
%% This is for dvips (IBM)
\includegraphics{diff.ps}
%% This is for specialhost (AMIGA)
%%\special{ifffile="diff.iff" hoffset=0pt voffset=0pt
%%         mode=bw hsize=150mm vsize=235mm}
%%\caption{Pionic transport coefficients}
\tenrm\baselineskip=12pt\noindent
Top panel: Thermal conductivity $\times$ temperature.\\
Bottom panel: Diffusion coefficient.\\[1mm]
Thin straight lines: slopes $\propto T^3$, $T^4$\\[1mm]
Full thick line: $\Delta$-hole model at $\rho$ = 1.59 nuclear density;\\
dashed line 0.92 and dash-dotted line 0.47 nuclear density.\\[1mm]
Dash-double-dotted in top panel \cite{W93}, dotted line
ref. \cite{G85},\\
 both from $\pi-\pi$ scattering data at $T=0$, $\rho=0$.
\end{figure}
%%%%%%%%%%%%%%%%%%%%%%%%%%%%%%%%%%%%%%%%%%%%%%%%%%%

The high temperature behavior at the example
density of $\rho =1.59\times$ nuclear matter density
inferred from the figure is
\begin{equation}
\lambda \approx 9897\,\mbox{GeV/fm$^2$}\,
\times\,\left[T/\mbox{GeV}\right]^{3.70}
\;,\end{equation}
i.e., it rises even faster than estimated by eqn. (\ref{asy}).
This is due to the decreasing width of the pion with temperature
in the $\Delta$-hole model -- a fact, which was
not included in eqn. (\ref{asy}). At lower temperatures,
the thermal conductivity rises slower with temperature than the
asymptotic expression.

The density dependence of the coefficients $\lambda$ and $d$
is quite small, with the general tendency to have a ''stiffer''
temperature dependence at lower baryon density.
This is consistent with the fact, that the pion in the $\Delta$-hole
model becomes a free particle without the baryonic background.

A comparison of these values to {\em experimental\/}  data is beyond
reach for the time being, we thus have to restrict the
comparison to other calculations:
The thermal conductivity in this field-theoretical calculation
is substantially higher, than obtained in a fluid-dynamical picture
fitted to pion scattering data at zero temperature
and density\cite{G85,W93}.

This is consistent with the finding, that {\em off-shell effects\/},
which were treated consistently in the present approach,
may substantially increase relaxation times \cite{D84a}.
Hence, within the framework of quantum field theory, the
new method presented here gives reliable results for
transport coefficients.
\clearpage
\section{Acknowledgements}
I would like to thank H.Umezawa for many discussions and important
remarks. To the organizing committee, I express my gratitude for
the support I received, and for the excellent conference organization.

%%%%%%%%%%%%%%%%%%%%%%%%%%%%%%%%%%%%%%%%%%%%%%%%%%%%%%%%%%%%%%%%%


\section{References}
\begin{thebibliography}{9}
\bibitem{L88}{
    N.P.Landsman, Phys.Rev.Lett. {\bf 60} (1988) 1990 and\\
    Ann.Phys. {\bf 186 } (1988) 141}
\bibitem{Ubook}{
    H.Umezawa,\\
    { Advanced Field Theory: Micro, Macro and Thermal Physics}\\
    (American Institute of Physics, 1993)}
\bibitem{SKM}{
    J.Schwinger,
    J.Math.Phys. {\bf 2} (1961) 407;\\
    L.V.Keldysh, Zh.Exsp.Teor.Fiz. {\bf 47} (1964) 1515 and
    JETP {\bf 20} (1965) 1018}
\bibitem{L86}{
    R.Littlejohn, Phys.Rep. {\bf 138} (1986) 193}
\bibitem{h90ber}{
    P.A.Henning, M.Graf and F. Matth\"aus,
    Physica {\bf A 182} (1992) 489}
\bibitem{WZ84}{
    F.Wilczek and A.Zee, Phys.Rev.Lett. {\bf 52} (1984) 2111}
\bibitem{hu92a}{
    P.A.Henning and H.Umezawa,
    Phys.Lett. {\bf B 303} (1993) 209}
\bibitem{habil}{
    P.A.Henning,\\
    {\em TFD for Quantum Fields with Continuous Mass Spectrum}\\
    Habilitationsschrift, October 1993\\
    to appear as GSI-Report}
\bibitem{h93pion}{
    P.A.Henning and H.Umezawa\\
    {\em The Delta-hole model at finite temperature}\\
     GSI Preprint 93-23 (1993),
     submitted to Nucl.Phys. {\bf A}}
\bibitem{NUY92}{
    K.Nakamura, H.Umezawa and Y.Yamanaka,\\
    Mod.Phys.Lett {\bf A7} (1992)  3583}
\bibitem{h93trans}{
    P.A.Henning \\
    {\em TFD and Kinetic Coefficients of a Charged Boson Gas},\\
    GSI-Preprint 93-45 (1993), Nucl.Phys. {\bf A} in press}
\bibitem{HST84}{
    A.Hosoya, M.Sakagami and M.Takao,
    Ann.Phys. {\bf 154} (1984) 229}
\bibitem{G85}{
    S.Gavin, Nucl.Phys. {\bf A435} (1985) 826}
\bibitem{W93}{
    M.Prakash, R.Venugopalan and G.Welke,\\
    {\em Non-equilibrium properties of hadronic mixtures},\\
    Wayne State University preprint (1993), to appear in Phys.Rep.}
\bibitem{D84a}{
    P.Danielewicz,
    Ann.Phys. {\bf 152} (1984)  239 and 305}
\end{thebibliography}
\end{document}